\documentclass[a4paper,11pt]{article}
\usepackage{pos}

\usepackage{subcaption}

\title{Lepton-flavour violation in hadronic tau decays\\ and \texorpdfstring{$\ell$--$\tau$}{l-tau} conversion in nuclei}
\ShortTitle{LFV in hadronic tau decays and \texorpdfstring{$\ell$--$\tau$}{l-tau} conversion in nuclei}

\makeatletter
\def\@fnsymbol#1{\ensuremath{\ifcase#1\or \dagger\or \ddagger\or \mathsection\or \mathparagraph\or \|\or **\or \dagger\dagger \or \ddagger\ddagger \else\@ctrerr\fi}}
\makeatother

\author*[a,\ddagger]{Tom\'a\v{s} Husek}
\author[b]{Kevin Mons\'alvez-Pozo}
\author[b]{Jorge Portol\'es}

\affiliation[a]{Institute of Particle and Nuclear Physics, Faculty of Mathematics and Physics, Charles University,\\
V Hole\v{s}ovi\v{c}k\'ach 2, 18000 Praha 8, Czech Republic%
\footnotemark[0]\footnotetext[0]{%
\hspace{-2.2mm}%
\setlength{\tabcolsep}{2pt}%
\begin{tabular}{rl}
$^\ddagger$Present address: & Department of Astronomy and Theoretical Physics, Lund University,\\
&S\"olvegatan 14A, SE 223-62 Lund, Sweden\end{tabular}}%
}

\affiliation[b]{Instituto de F\'isica Corpuscular, CSIC -- Universitat de Val\`encia,\\
Apt.\ Correus 22085, E-46071 Val\`encia, Spain}

\emailAdd{tomas.husek@thep.lu.se}
\emailAdd{kevin.monsalvez@ific.uv.es}
\emailAdd{jorge.portoles@ific.uv.es}

\abstract{
Within the Standard Model Effective Field Theory framework, with operators up to dimension~6, we perform a model-independent analysis of the lepton-flavour-violating processes involving tau leptons.
Namely, we study hadronic tau decays and $\ell$--$\tau$ conversion in nuclei ($\ell=e,\mu$).
Based on available experimental limits, we establish constraints on the Wilson coefficients of the operators contributing to these processes.
The related information from Belle II and foreseen future experiments can be easily incorporated into the resulting framework.
}

\FullConference{%
  40th International Conference on High Energy physics - ICHEP2020\\
  July 28 - August 6, 2020\\
  Prague, Czech Republic (virtual meeting)
}

\begin{document}
\maketitle

\section{Introduction}

The lepton sector of the Standard Model is not particularly rich in flavour phenomena, at least as compared to the quark sector.
Even though neutral leptons---neutrinos---oscillate, flavour violation in the charged-lepton sector has not been observed: Indeed, the minimal extension of the Standard Model predicts that these processes are very suppressed.
On the other hand, new-physics scenarios allow for enhanced charged-lepton-flavour violation (CLFV).
We shall try to collect all the available information from various types of experiments and constrain these scenarios in a systematic way.

In the literature to date, CLFV processes involving mainly muons or electrons were discussed.
In the recent work~\cite{Husek:2020fru}, we look into processes involving tau leptons.
This third-generation charged lepton comes with a unique feature as it also allows us to study hadronic ($\tau$) decays.
We thus analyze $\tau$-involved processes which are connected to past experiments, as for instance Belle (thus there are already existing limits)~\cite{Amhis:2019ckw}.
Moreover, we take into account limits expected by Belle II~~\cite{Kou:2018nap}: It is anticipated that these limits on hadronic tau decays will be improved by roughly one order of magnitude.
As another input to our analysis, we also consider expected sensitivities of the NA64 experiment at CERN~\cite{Gninenko:2018num} for the cross section of the $\ell$--$\tau$ conversion in nuclei.

\section{Calculation}

In the Standard Model Effective Field Theory (SMEFT) framework, we calculated the following three sets of hadronic tau decays: $\tau^- \rightarrow \ell^- P$, $\tau^- \rightarrow \ell^- P_1P_2$ and $\tau^- \rightarrow \ell^- V$ ($P$ stands for any pseudoscalar meson, $V$ for a vector resonance), as well as the $\ell$--$\tau$ conversion ($\ell=e,\mu$) in Fe(56,26) and Pb(208,82), and combined all the above in a global numerical analysis.
The SMEFT allows us to parameterize, at the electroweak scale, effects that might appear due to new physics emerging at scale $\Lambda$, in a model-independent and systematic way:
\begin{equation}
{\mathcal{L}_{{\text{SMEFT}}}}
={\mathcal{L}}_{{\text{SM}}}
+\sum_{D>4}\frac{1}{\Lambda^{{\text{D}}-4}} \sum_i \, C_i^{{\text{(D)}}} \, {\cal O}^{{\text{(D)}}}_i\,.
\label{eq:smeft}
\end{equation}
In this analysis, we take into account CLFV operators up to $D=6$
based on Ref.
~\cite{Grzadkowski:2010es} and listed in Table~\ref{tab:1}.
\begin{table}[!tb]
\begin{center}
\renewcommand{\arraystretch}{1.2}
\scalebox{0.85}{
\begin{tabular}{|c|c||c|c|}
\hline
WC & Operator & WC & Operator\\
\hline
\hline
 $C_{LQ}^{(1)}$ & $\left( \bar{L}_p \gamma_{\mu} L_r \right) \left( \bar{Q}_s \gamma^{\mu} Q_t \right)$ & $C_{e \varphi}$  &
 $\left( \varphi^{\dagger} \varphi \right) \left( \bar{L}_p e_r \varphi \right)$   \\
 $C_{LQ}^{(3)}$ & $\left( \bar{L}_p \gamma_{\mu} \sigma^I L_r \right) \left( \bar{Q}_s \gamma^{\mu} \sigma^I Q_t \right)$ &
 $C_{\varphi e}$  &
 $\left( \varphi^{\dagger}  i \overset{\leftrightarrow}{D}_{\mu} \varphi \right) \left( e_p \gamma^{\mu} e_r  \right)$   \\
 $C_{eu}$ & $\left( \bar{e}_p \gamma_{\mu} e_r \right) \left( \bar{u}_s \gamma^{\mu} u_t \right)$ & $C_{\varphi L}^{(1)}$ &
 $\left( \varphi^{\dagger} i \overset{\leftrightarrow}{D}_{\mu} \varphi \right) \left( \bar{L}_p \gamma^{\mu} L_r \right)$ \\
 $C_{ed}$ &  $\left( \bar{e}_p \gamma_{\mu} e_r \right) \left( \bar{d}_s \gamma^{\mu} d_t \right)$ & $C_{\varphi L}^{(3)}$ &
 $\left( \varphi^{\dagger} i \overset{\leftrightarrow}{D}_{I\mu} \varphi \right) \left( \bar{L}_p \sigma_I \gamma^{\mu} L_r \right)$ \\
 $C_{Lu}$ & $\left( \bar{L}_p \gamma_{\mu} L_r \right) \left( \bar{u}_s \gamma^{\mu} u_t \right)$ & $C_{eW}$ & $ \left( \bar{L}_p \sigma^{\mu \nu} e_r \right) \sigma_I \varphi W^I_{\mu \nu}$ \\
 $C_{Ld}$ &  $\left( \bar{L}_p \gamma_{\mu} L_r \right) \left( \bar{d}_s \gamma^{\mu} d_t \right)$ & $C_{eB}$ & $ \left( \bar{L}_p \sigma^{\mu \nu} e_r \right) \varphi B_{\mu \nu}$ \\ 
 $C_{Qe}$ &  $\left( \bar{Q}_p \gamma_{\mu} Q_r \right) \left( \bar{e}_s \gamma^{\mu} e_t \right)$ & & \\
 $C_{LedQ}$ & $\left( \bar{L}^j_p e_r \right) \left( \bar{d}_s Q^j_t \right)$ & & \\ 
 $C_{LeQu}^{(1)}$ & $\left( \bar{L}_p^j e_r \right) \varepsilon_{jk}  \left( \bar{Q}_s^k u_t \right)$ & & \\
 $C_{LeQu}^{(3)}$ & $\left( \bar{L}_p^j \sigma_{\mu \nu} e_r \right)  \varepsilon_{jk}  \left( \bar{Q}_s^k \sigma^{\mu \nu} u_t \right)$ & & \\
\hline
\end{tabular}
}
\end{center}
\caption{\label{tab:1}
$D=6$ operators appearing in the Lagrangian (\ref{eq:smeft}) and contributing to the CLFV processes that we study in Ref.~\cite{Husek:2020fru}. The four-fermion operators appear on the left-hand side, while those involving the Higgs doublet $\varphi$ and the gauge bosons are on the right. The notation (up to small apparent changes) follows Ref.~\cite{Grzadkowski:2010es}.}
\end{table}
Our aim is then to put limits on the corresponding Wilson coefficients (WCs), which, in this setting, are expected to be of order one.

\begin{figure}[!b]
\vspace{-6mm}
\begin{center}
\subfloat[][]{
\includegraphics[width=0.21\textwidth]{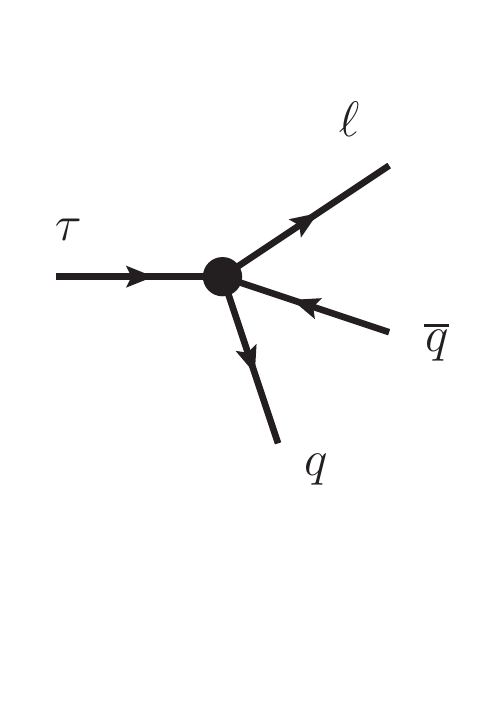}\vspace{-8mm}
}
\subfloat[][]{
\includegraphics[width=0.21\textwidth]{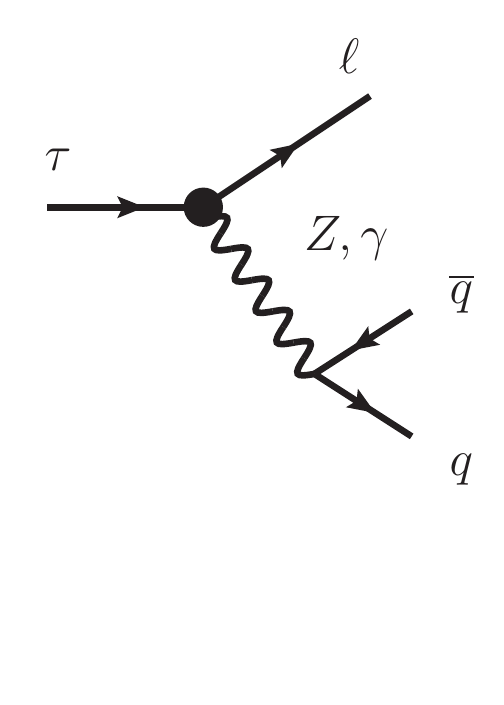}\vspace{-10mm}
}
\subfloat[][]{
\includegraphics[width=0.21\textwidth]{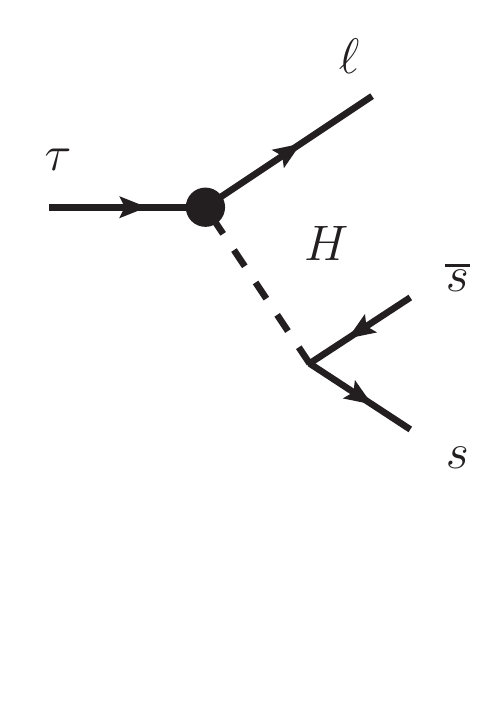}\vspace{-10mm}
}
\subfloat[][]{
\includegraphics[width=0.32\textwidth]{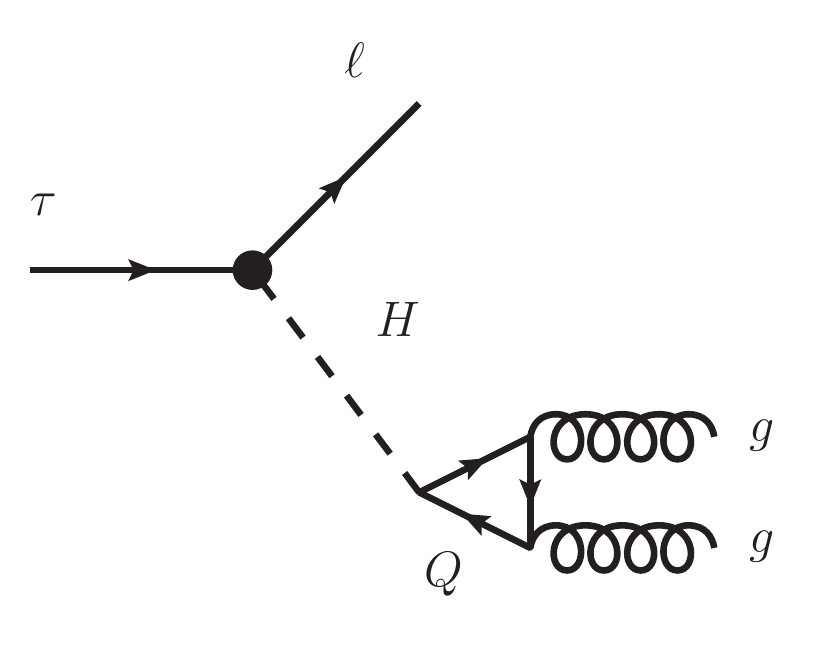}
}
\end{center}
\vspace{-5mm}
\caption[]{\label{fig:1} Contributions of the SMEFT Lagrangian to $\tau \rightarrow \ell \, \overline{q} q$ ((a)-(c)), and the dominant scalar contribution to $\tau \rightarrow \ell \overline{P}P$ ((d)), with $\ell = e,\mu$ and $P=\pi, K$.
The dot indicates the CLFV vertex.
We consider $m_u=m_d=0$ and $m_s \neq 0$; $Q$ in the loop stands for a heavy quark: $Q=c,b,t$.}
\end{figure}

\begin{figure}[!tb]
\vspace{-4mm}
\begin{center}
\subfloat[][]{
\includegraphics[width=0.155\textwidth]{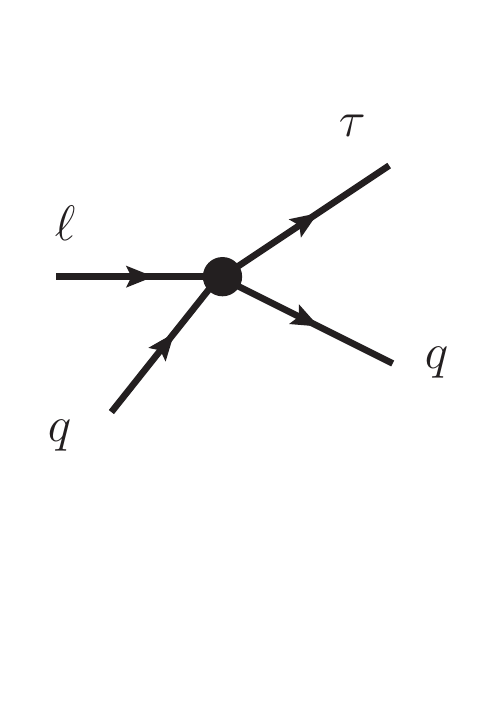}\vspace{-9mm}
}
\subfloat[][]{
\includegraphics[width=0.155\textwidth]{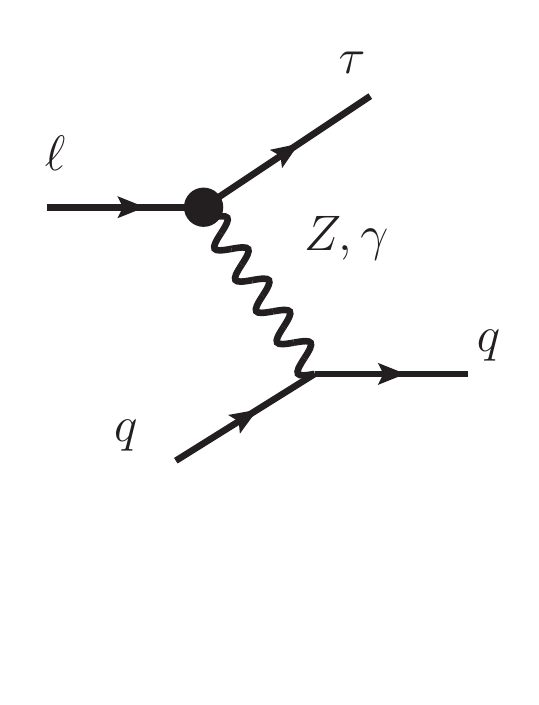}\vspace{-7mm}
}
\subfloat[][]{
\includegraphics[width=0.155\textwidth]{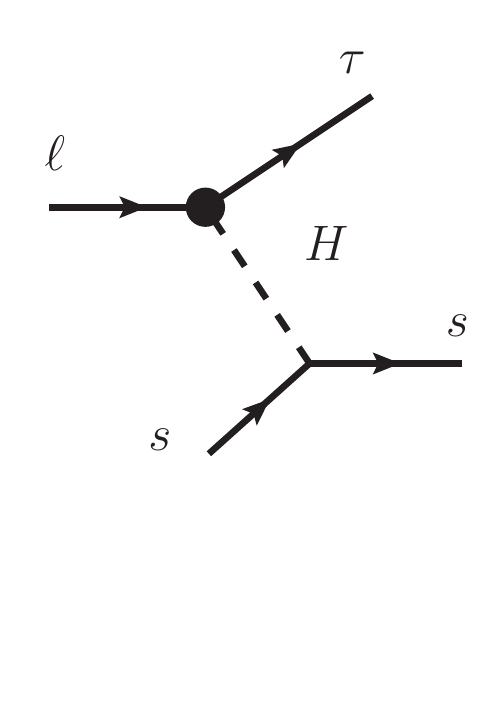}\vspace{-8mm}
}
\subfloat[][]{
\includegraphics[width=0.24\textwidth]{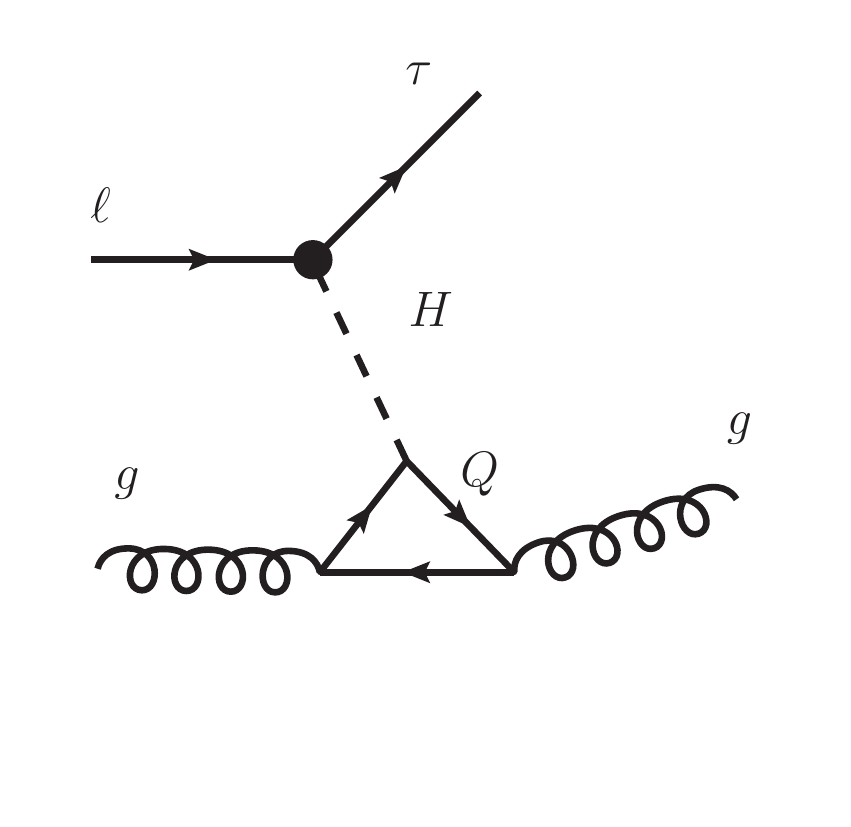}\vspace{-8mm}
}
\subfloat[][]{
\includegraphics[width=0.24\textwidth]{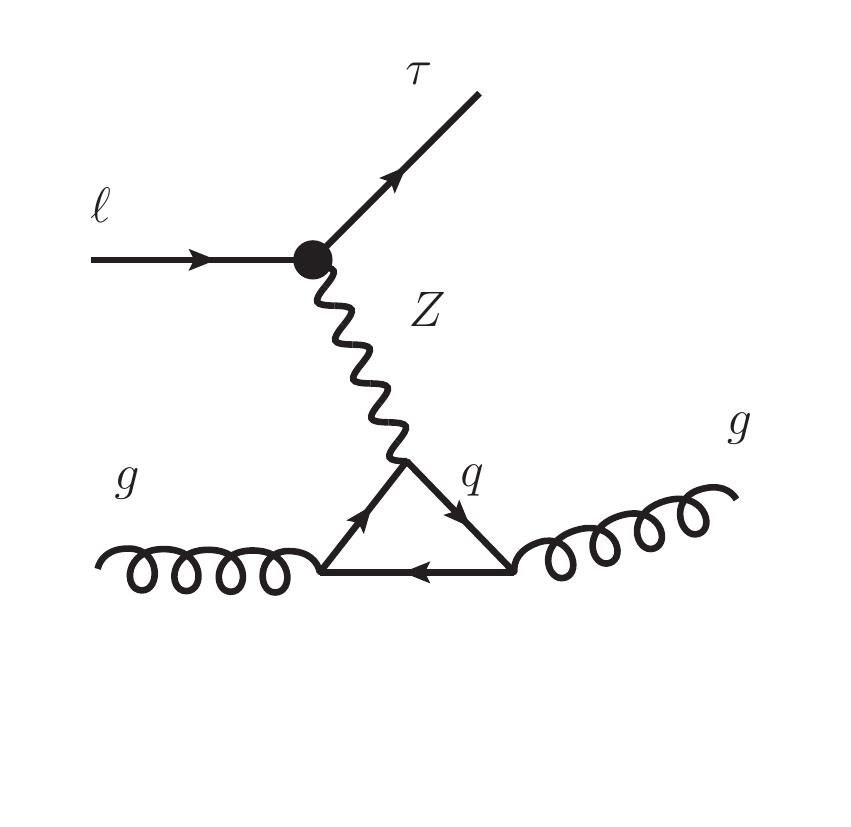}\vspace{-8mm}
}
\caption[]{\label{fig:3} Contributions of the SMEFT Lagrangian to $\ell q \rightarrow \tau q$, with $q=u,d,s$ ((a)-(c)), and Higgs and $Z$ contribution to $\ell g \rightarrow \tau g$, with $Q=c,b,t$ and $q=u,d,c,s,t,b$ ((d), (e)).
We put $m_u=m_d=0$; $m_s \neq 0$.}
\end{center}
\end{figure}

Regarding the hadronic $\tau$ decays, the diagrams contributing to the perturbative part of the amplitude are shown in Figure~\ref{fig:1}.
Besides the contributions containing the quark--antiquark pair in the final state (as typically considered in literature), we also included two-gluon state, represented by the dominant Higgs-exchange contribution~\cite{Celis:2013xja}.
The quark currents that appear in the final state need to be hadronized, which we do in terms of the Chiral Perturbation Theory and Resonance Chiral Theory~\cite{Ecker:1988te}, the latter of which allows us to introduce resonances as explicit degrees of freedom.

For the $\mu$--$\tau$ conversion in nuclei, contributions to the perturbative part of the amplitude are shown in Figure~\ref{fig:3}.
As before, the incident lepton can either interact with a quark in the nucleus, or analogically with an anti-quark, and also with a gluon.
The last two diagrams in Figure~\ref{fig:3} represent the dominant contributions for the gluon part of the cross section.
The above mentioned partons are confined inside the nuclei, so there are low-energy non-perturbative QCD effects present.
Once the calculation of the perturbative part of the amplitude is performed, to obtain the total cross section the results are convoluted with nuclear parton-distribution functions (nPDFs).
For the latter, we use the fit done by the group around the nCTEQ15 project~\cite{Kovarik:2015cma}.

Some of the Wilson coefficients always contribute to the processes that we consider in the same way. We are thus not able to fit these WCs separately, but only some of their combinations.
We hence redefine the set of variables we fit.
Moreover, in the contact interaction with flavour-changing neutral currents (diagrams (a) in Figures~\ref{fig:1} and \ref{fig:3}), we allow for quarks to change flavour, although considering the same Wilson coefficients for all the quark flavours.

In the end, we thus have a set of hadronic observables with existing or expected bounds, each of them depending on several WCs.
In total, we end up with 15 free parameters entering the global numerical analysis.
Subsequently, we use $\texttt{HEPfit}$~\cite{deBlas:2019okz}, an open-source tool based on Bayesian statistical framework, to fit the WCs-over-$\Lambda^2$ ratios.
It allows us to sample the whole parameter space of WCs and gives us the range of allowed values with different confidence levels.
It is also important to mention the priors: We assume flat initial distributions for the WCs.

\section{Results}

\begin{figure}[!tb]
\begin{center}
\vspace{-2mm}
\includegraphics[width=0.8\textwidth]{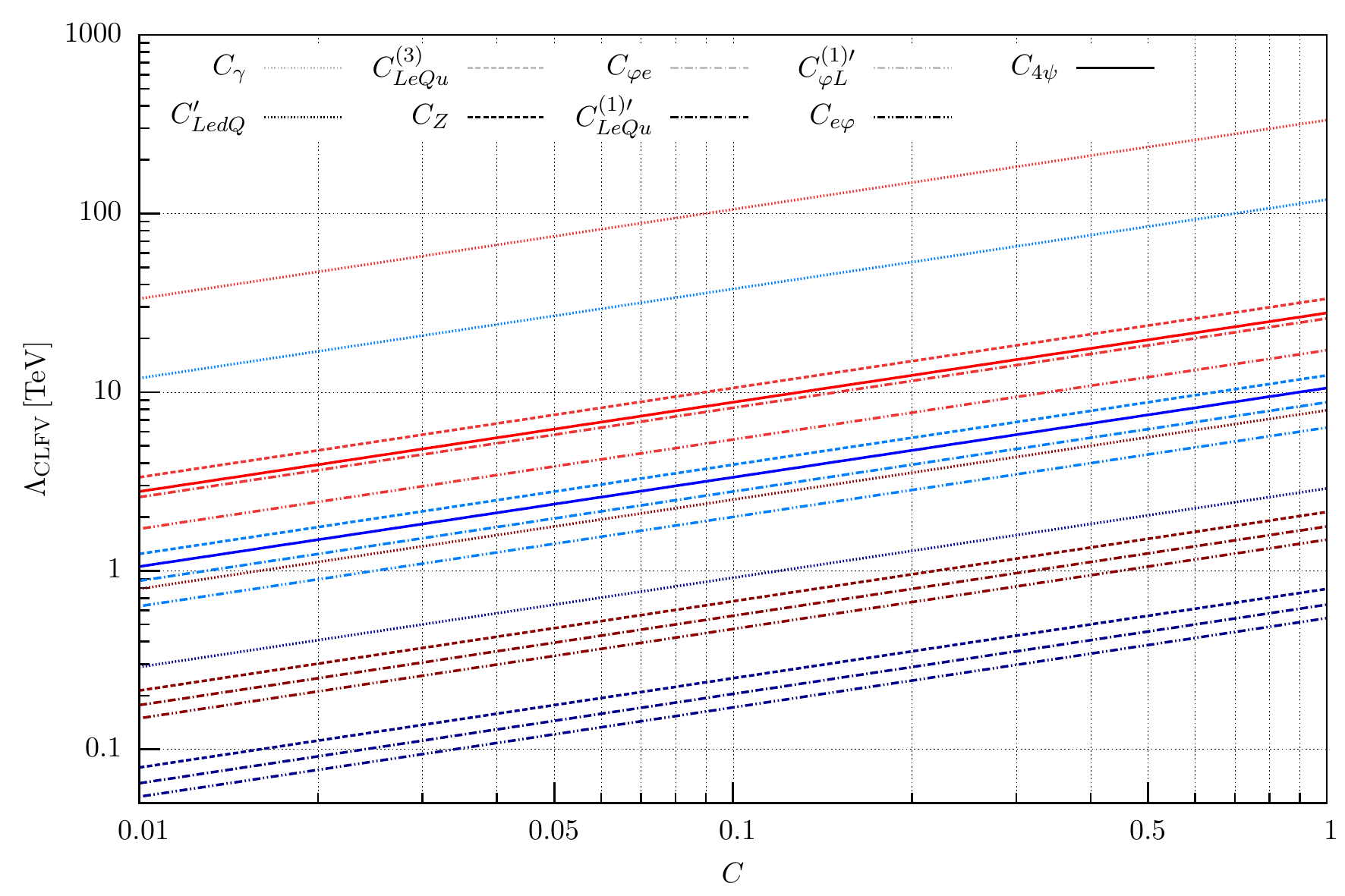}
\caption[]{\label{fig:5} Constraints on $\Lambda_\text{CLFV}$ with respect to the values of WCs, based on the current Belle (shades of blue) and expected Belle II (shades of red) limits, given at the 99.8\,\% confidence level.
The four-fermion WCs are represented altogether as $C_{4\psi}$.
For a given set of limits (color), the lighter shades correspond to the WCs listed in the first row of the key (omitting $C_{4\psi}$) and the darker shades to those from the second row.}
\end{center}
\end{figure}

\begin{figure}[!tb]
\begin{center}
\vspace{-2mm}
\includegraphics[width=0.85\textwidth]{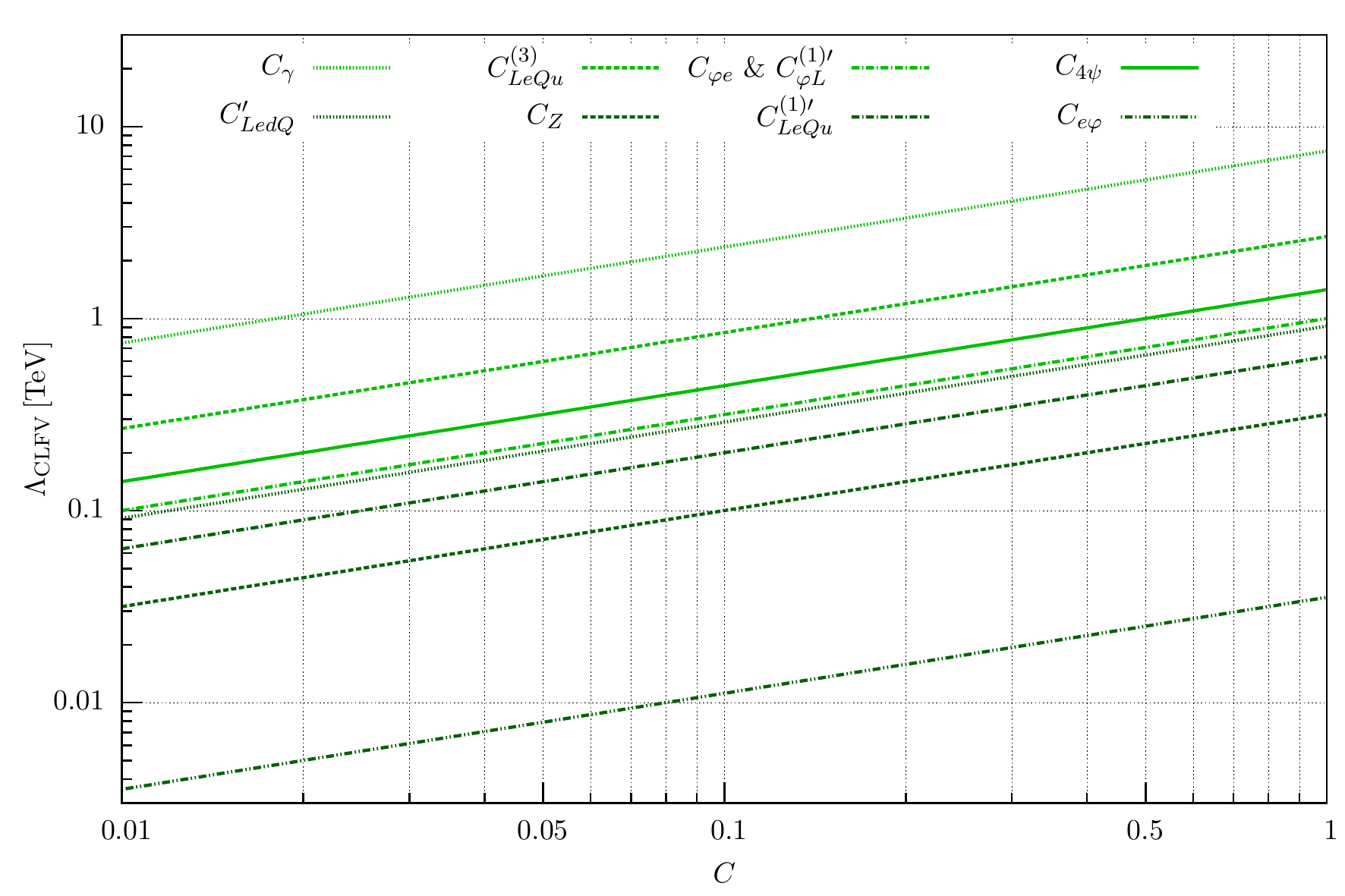}
\caption[]{\label{fig:7} Constraints on $\Lambda_{{\text{CLFV}}}$ with respect to the values of WCs from $\mu$-$\tau$ conversion in Fe(56,26), based on the expected sensitivity of the NA64 experiment, given at the 99.8\,\% confidence level.}
\end{center}
\end{figure}

\begin{figure}[!tb]
\begin{center}
\vspace{-2mm}
\includegraphics[width=0.77\textwidth]{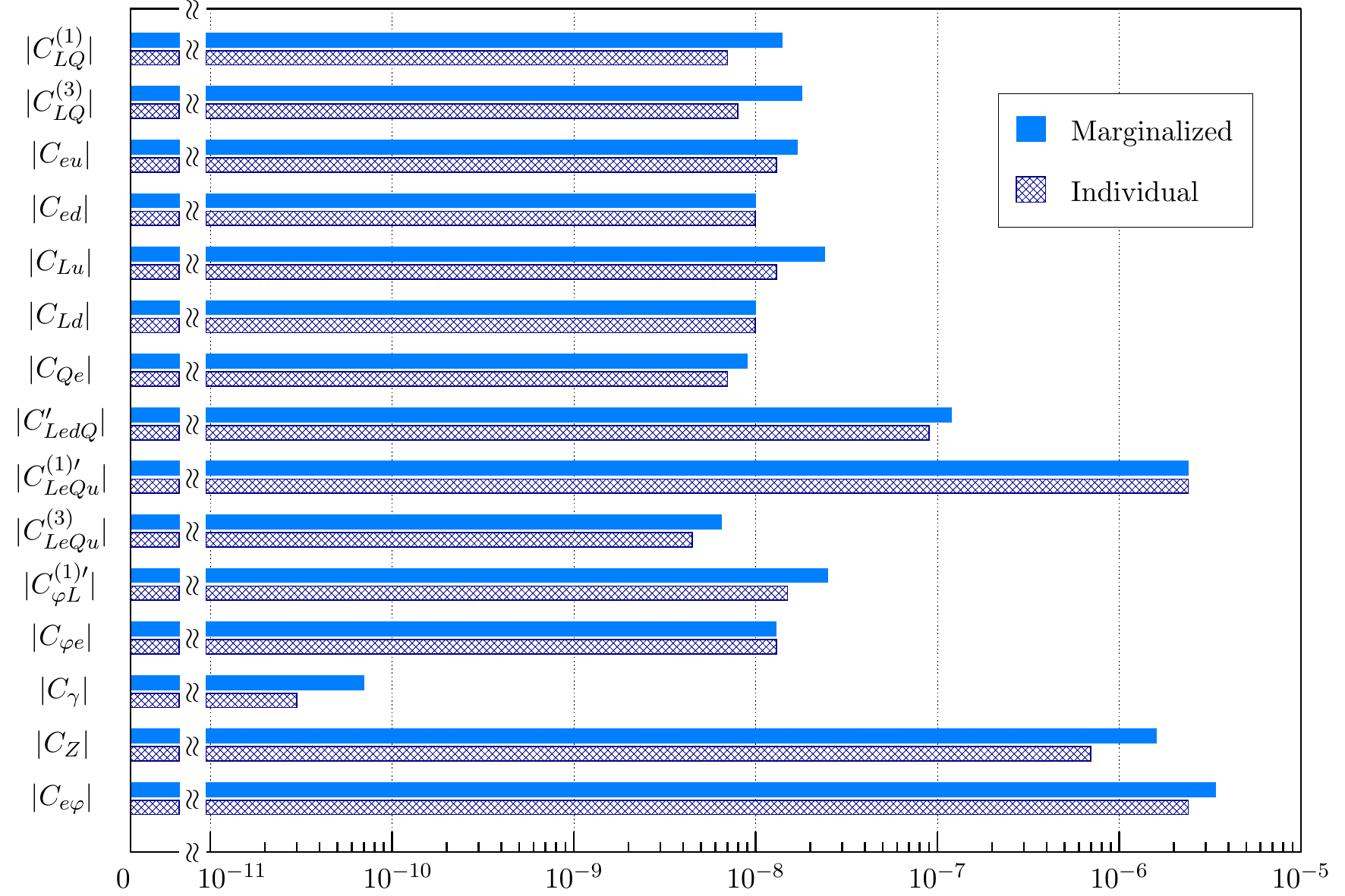}
\caption[]{\label{fig:6} Constraints on $C/ \Lambda_{{\text{CLFV}}}^{2}$\,[GeV$^{-2}$] based on the current Belle limits, stemming from the marginalized and individual analyses for tau decays, given at the 99.8\% confidence level.}
\end{center}
\end{figure}

The results regarding the hadronic tau decays are shown in Figure~\ref{fig:5}.
%
Assuming that the coefficients would be of order 1, one can extract the information on the scale where the physics related to the CLFV phenomena emerges.
This translates into Table~\ref{tab:Belle} which shows, for instance, that if only the coefficient $C_\gamma$ would be responsible for new physics, we could conclude that, based on the current limits, the related scale lies above $\approx$120\,TeV.
\begin{table}[!tb]
\begin{subtable}[t]{0.48\textwidth}
\begin{center}
\renewcommand{\arraystretch}{1.5}
\scalebox{0.75}{
\begin{tabular}{||c|c|c|c|c|c|}
\hline
 WC & Belle & Belle II & WC &  Belle & Belle II\\
\hline
\hline
 $C_{LQ}^{(1)}$ & $\gtrsim 8.5 $  & $\gtrsim 26 $  &  $C_{LeQu}^{(1) \, \prime}$ & $\gtrsim 0.65 $  & $\gtrsim 1.8 $  \\
\hline
 $C_{LQ}^{(3)}$ & $\gtrsim 7.5 $   & $\gtrsim 21 $ &  $C_{LeQu}^{(3)}$ & $\gtrsim 12$ & $\gtrsim 33 $ \\
\hline
 $C_{eu}$ &  $\gtrsim 7.7 $  & $\gtrsim 22 $ & $C_{\varphi L}^{(1) \, \prime}$ & $\gtrsim 6.3 $  & $\gtrsim 17 $  \\
\hline
$C_{ed}$, $C_{Ld}$ & $\gtrsim 10 $   & $\gtrsim 26 $  & $C_{\varphi e}$ & $\gtrsim 8.8 $  & $\gtrsim 26 $  \\
\hline
$C_{Lu}$ & $\gtrsim 6.5 $  & $\gtrsim 20 $   & $C_{\gamma}$ & $\gtrsim 120$  & $\gtrsim 330$ \\
\hline
$C_{Qe}$ & $\gtrsim 11 $  & $\gtrsim 28 $  & $C_{Z}$ & $\gtrsim 0.79$  & $\gtrsim 2.1 $ \\
\hline
$C_{LedQ}^{\, \prime}$ & $\gtrsim 2.9 $  & $\gtrsim 7.9 $  &  $C_{e \varphi }$ & $\gtrsim 0.54$  & $\gtrsim 1.5 $ \\
\hline
\end{tabular}
}
\end{center}
\vspace*{-0.5cm}
\caption{\label{tab:Belle} Tau decays (based on Belle and expected Belle II limits).}
\end{subtable}
\hfill
\begin{subtable}[t]{0.48\textwidth}
\begin{center}
\renewcommand{\arraystretch}{1.5}
\scalebox{0.75}{
\begin{tabular}{||c|c|c|c|c|c|}
\hline
 WC & $e$--$\tau$ & $\mu$--$\tau$ & WC &  $e$--$\tau$ & $\mu$--$\tau$  \\
\hline
\hline
 $C_{LQ}^{(1)}$ & $\gtrsim 0.13 $  & $\gtrsim 1.7 $  &  $C_{LedQ}$ & $\gtrsim 0.06$  & $\gtrsim 0.9  $  \\
\hline
 $C_{LQ}^{(3)}$ & $\gtrsim 0.11 $   & $\gtrsim 1.5 $ &  $C_{LeQu}^{(1)}$ & $\gtrsim 0.05$ & $\gtrsim 0.6 $ \\
\hline
 $C_{eu}$ &  $\gtrsim 0.11 $  & $\gtrsim 1.4 $ & $C_{LeQu}^{(3)}$ & $\gtrsim 0.2 $  & $\gtrsim 2.7 $  \\
\hline
$C_{ed}$ & $\gtrsim 0.11 $   & $\gtrsim 1.4 $  & $C_{\varphi e}, C_{\varphi L}^{(1)}$ & $\gtrsim 0.08 $  & $\gtrsim 1 $  \\
\hline
$C_{Lu}$ & $\gtrsim 0.09 $  & $\gtrsim 1.1 $   & $C_{\gamma}$ & $\gtrsim 0.6$  & $\gtrsim 7.5 $ \\
\hline
$C_{Ld}$ & $\gtrsim 0.09 $  & $\gtrsim 1.2 $  & $C_{Z}$ & $\gtrsim 0.02$  & $\gtrsim 0.3 $ \\
\hline
$ C_{Qe} $ & $\gtrsim 0.1 $  & $\gtrsim 1.4 $  &  $C_{e \varphi }$ & $\gtrsim 0.003$  & $\gtrsim 0.04 $ \\
\hline
\end{tabular}
}
\end{center}
\vspace*{-0.5cm}
\caption{\label{tab:ell-tau} $\ell$--$\tau$ conversion in Fe(56,26) (based on the expected sensitivity of the NA64 experiment).}
\end{subtable}
\caption{Bounds on the new-physics energy scale mediating CLFV phenomena ($\Lambda_{{\text{CLFV}}}$) given in TeV. Here, we consider $C\approx1$. The results are given at the 99.8\,\% confidence level.}
\end{table}
On the other hand, there are coefficients with much weaker bounds.
The analysis also shows that some of the WCs are rather correlated, which then translates to the fact that, if we perform individual analyses (i.e.\ if we set only one of the WCs non-zero), we would gain much stronger bounds than in the case when we allow for all the coefficients to vary (see Fig.~\ref{fig:6}).
If there are correlations among the coefficients, the sensitivity is diluted among the correlated WCs.
In other words, we obtain more conservative results when assuming all these coefficients to be non-zero simultaneously.
For completeness, Figure~\ref{fig:7} shows the results for the $\mu$--$\tau$ conversion in Fe(56,26) based on the expected NA64 sensitivity which cannot currently compete with the Belle (II) limits:
Another improvement of at least two orders of magnitude would be necessary, which in turn would provide valuable complementary inputs to resolve some of the correlations among WCs.
The limits are much weaker compared to the hadronic tau decays.
This translates into the values listed in Table~\ref{tab:ell-tau}.

\section{Summary}

We performed model-independent numerical analysis of the SMEFT dimension-6 operators related to CLFV processes involving the $\tau$ lepton.
In particular, we studied 28 hadronic $\tau$ decay channels and four $\ell$--$\tau$ conversion cross-sections in Fe(56,26) and Pb(208,82), out of which the constraints imposed by the $\mu$--$\tau$ conversion in iron are the strongest.
As for the experimental inputs, we used present Belle as well as expected Belle II limits, and expected sensitivities of NA64 for $\ell$--$\tau$ conversion in nuclei.
The $\texttt{HEPfit}$ tool was used for the statistical part of the project.

\section{Acknowledgement}

This work has been supported in part by Grants No.\ FPA2017-84445-P and SEV-2014-0398 (AEI/ERDF, EU), by PROMETEO/2017/053 (GV), by Czech Science Foundation grant GA\v{C}R 18-17224S, and by Swedish Research Council grants contract numbers 2016-05996 and 2019-03779.


\providecommand{\href}[2]{#2}\begingroup\raggedright\endgroup

\end{document}